\documentclass[american,aps,pra,reprint, superscriptaddress]{revtex4-1}
\usepackage{amsmath,amscd,amsthm,amssymb}
\usepackage{mathrsfs}
\usepackage{graphicx}
\usepackage{epsf,epstopdf}
\usepackage[center]{caption2}
\usepackage[all]{xy}
\usepackage[unicode=true,pdfusetitle, bookmarks=true,bookmarksnumbered=false,bookmarksopen=false, breaklinks=false,pdfborder={0 0 0},backref=false,colorlinks=false] {hyperref}
\hypersetup{ colorlinks,linkcolor=myurlcolor,citecolor=myurlcolor,urlcolor=myurlcolor}
\usepackage{graphics,epstopdf,graphicx, amsthm, amsmath, amssymb, times, braket, color, bm}
\usepackage[up]{subfigure}
\usepackage{cleveref}

\definecolor{myurlcolor}{rgb}{0,0,0.7}

\theoremstyle{plain}
\newtheorem{thm}{\protect\theoremname}
\newtheorem{prop}[thm]{Proposition}
\newtheorem{lem}[thm]{Lemma}
\providecommand{\theoremname}{Theorem}
\newcommand*{\myproofname}{Proof}

\makeatother

\newtheorem*{cor}{Corollary}

\theoremstyle{definition}

\theoremstyle{remark}
\newtheorem{rem}{Remark}

\begin{document}

 \author{Chunhe Xiong}
 \email{xiongchunhe@zju.edu.cn}
 \affiliation{School of Mathematical Sciences, Zhejiang University, Hangzhou 310027, PR~China}

 \author{Junde Wu}
 \email{corresponding author: wjd@zju.edu.cn}
 \affiliation{School of Mathematical Sciences, Zhejiang University, Hangzhou 310027, PR~China}

\title{Geometric coherence and quantum state discrimination}
\begin{abstract}

 The operational meaning of coherence measure lies at very heart of the coherence theory. In this paper, we provide an operational interpretation for geometric coherence, by proving that the geometric coherence of a quantum state is equal to the minimum error probability to discriminate a set of pure states with von Neumann measurement. On the other hand, we also show that a task to ambiguously discriminate a set of linearly independent pure states can be also regards as a problem of calculating geometric coherence. That is, we reveal an equivalence relation between ambiguous quantum state discrimination and geometric coherence. Based on this equivalence, moreover, we improve the upper bound for geometric coherence and give the explicit expression of geometric coherence for a class of states. Besides, we establish a complementarity relation of geometric coherence and path distinguishability, with which the relationship between $l_1$-norm of coherence and geometric coherence is obtained. Finally, with geometric coherence, we study multiple copies quantum state discrimination and give an example to show how to discriminate two pure states.

\end{abstract}
\maketitle

\section{introduction}

 Coherence is a fundamental feature of quantum mechanics, characterizing the wave nature for all objects. It is an essential correlation in quantum information theory \cite{Adesso2014A,Yao2015}, and plays an important role in quantum biology \cite{Plenio2008,Huelga13}, metrology \cite{Giovannetti2011} and thermodynamics \cite{Lostaglio2015A,Lostaglio2015B}. Quantum coherence also can be regarded as a kind of resource and its role in quantum algorithm has been investigated in \cite{Matera2016A,Hillery2016A,Anand2017,Shi2017}. The research on coherence in optics has lasted for a long time \cite{Glauber63,Sudarshan63} and only in recent years there develops a theory to quantify coherence \cite{Baumgratz2014}.

As well as entanglement \cite{HorodeckiRMP09}, \citeauthor{Baumgratz2014} provide a resource theory framework for coherence and a great deal of efforts have been done to further develop this theory \cite{Girolami14,Streltsov2015B,Yuan2015,winter2016,Chitambar2016A,chitambar2016B,streltsov2017A,Bu2017A}. In this frame, coherence quantifies the superpositions of a state in a fixed standard orthogonal basis. Given an orthonormal basis $\{\ket{i}\}^d_{i=1}$ for a $d$-dimensional Hilbert space $\mathcal{H}$, density matrices that are diagonal in this basis are called incoherent states and we label this set of quantum states by $\mathcal{I}$. Hence, a density matrix $\sigma\in\mathcal{I}$ is of the form
\begin{align}
\sigma=\sum^d_{i=1}\lambda_i\ket{i}\bra{i}.
\end{align}

 Any trace preserving completely positive (CPTP) map is called incoherent operation (IO) if one of its Kraus representation $\{K_i\}$ is incoherent. That is, each $K_i$ satisfies
\begin{align}
\frac{K_i\rho K^{\dagger}_i}{\mathrm{Tr}K_i\rho K^{\dagger}_i}\in\mathcal{I},
\end{align}
for any incoherent state $\rho$.

Similar to the quantification of entanglement \cite{HorodeckiRMP09,Vedral1997,Vedral1998,Plenio2000}, the authors in \cite{Baumgratz2014} propose the following conditions to be satisfied as a measure of coherence $C(\rho)$:

(C1) Faithful: $C(\rho)\ge0$ with equality if and only if $\rho$ is incoherent.

(C2) Monotonicity: $C$ does not increase under the action of incoherent operation, i.e., $C(\Phi(\rho))\le C(\rho)$ for any incoherent operation $\Phi$.

(C3) Strong monotonicity: $C$ does not increase on average under selective incoherent operations, i.e., $\sum_ip_iC(\sigma_i)\le C(\rho)$ with probabilities $p_i=\mathrm{Tr}K_i\rho K^{\dagger}_i$, post-measurement states $\sigma_i=p^{-1}_iK_i\rho K^{\dagger}_i$, and incoherent operators $K_i$.

(C4) Convexity: Nonincreasing under mixing of quantum states, i.e.,$\sum_ip_iC(\rho_i)\ge C(\sum_ip_i\rho_i))$ for any states $\{\rho_i\}$ and $p_i\ge0$ with $\sum_ip_i=1$.

 Analogous with geometric entanglement \cite{wei2003,Streltsov2010} which is an entanglement measure, the authors in \cite{Streltsov2015B} define geometric coherence as
\begin{align}
C_g(\rho)=1-F(\rho),\nonumber
\end{align}
where $F(\rho):=\mathop{\mathrm{max}}_{\sigma\in\mathcal{I}}F(\rho,\sigma)$ and the fidelity $F$ is defined as
\begin{align}
F(\rho,\sigma)=(\mathrm{Tr}\sqrt{\sqrt{\sigma}\rho\sqrt{\sigma}})^2.\nonumber
\end{align}

We call an incoherent state $\sigma_{\rho}$ the closest incoherent state (CIS) of $\rho$ if $F(\rho)=F(\rho,\sigma_{\rho})$.

In \cite{Streltsov2015B}, the authors show that geometric coherence is a coherence measure. In this paper, we provide an operational interpretation for $C_g$ by linking the geometric coherence with the task of quantum state discrimination (QSD). As a fundamental problem in quantum mechanics, QSD has been the subject of active theoretical investigation for a long time \cite{Helstrom1976,Holevo2011,Holevo2001,HELSTROM1967A,HELSTROM1968,HOLEVO1973337,Yuen1975,Davies1978}. Besides, QSD also plays an important role in quantum communication and quantum cryptography \cite{Phoenix1995,Bouwmeester2000,Gisin2002RMP,Loepp2006}.

In the task of ambiguous QSD, the sender draws at random some states from $\rho_i$'s with probabilities $\eta_i$ and send them to the receiver, whose job is to determine which state he has received as accurate as possible. To do so, he performs a positive operator valued measure (POVM) on each $\rho_i$ and declares the state is $\rho_j$ when the measurement reads $j$. The POVM is a set of positive operator $\{M_i\}$ satisfying $M_i=I$. As the probability to get the result $j$ is $p_{j|i}=\mathrm{Tr}(M_j\rho_i)$ provided the system is in the state $\rho_i$, the maximal success probability to identify $\{\rho_i,\eta_i\}$ is
\begin{align}
P^{opt}_S(\{\rho_i,\eta_i\})=\mathop{\mathrm{max}}_{\{M_i\}}\sum_i\eta_i\mathrm{Tr}(M_i\rho_i),\nonumber
\end{align}
where the maximum are over all POVM $\{M_i\}$ and moreover, the minimal error probability is
\begin{align}
P^{opt}_E(\{\rho_i,\eta_i\})=1-\mathop{\mathrm{max}}_{\{M_i\}}\sum_i\eta_i\mathrm{Tr}(M_i\rho_i).\nonumber
\end{align}

For two states, the analytic formula of maximal success probability $P^{opt}_S$ and the corresponding optimal measurement has been known, however, no solution is known for general case except some symmetric cases \cite{Bergou2004,Eldar2004,Chou2003}.

In this paper, we reveal the equivalence relation between geometric coherence and ambiguous QSD. Based on the equivalence, we will get some results of both geometric coherence and QSD.

The paper is organised as follows. In section \uppercase\expandafter{\romannumeral2} and section \uppercase\expandafter{\romannumeral3} we reveal the equivalence between geometric coherence and ambiguous QSD. We establish a complementarity relationship of geometric coherence and path distinguishability in \uppercase\expandafter{\romannumeral4}. Based on the equivalence, we obtain some results of both in section \uppercase\expandafter{\romannumeral5} and section \uppercase\expandafter{\romannumeral6}. Besides, We conclude in Section \uppercase\expandafter{\romannumeral7} with a summary and outlook.

\section{geometric coherence and quantum state discrimination}

It is well known that POVM can perform better than von Neumann measurement in quantum state discrimination\cite{Peres1990,peres1991}. However, the optimal measurement to discriminate a collection of linearly independent states, both pure \cite{kennedy1973} and mixed \cite{Eldar2003}, is still the von Neumann measurement. With this result, we can establish the relation between geometric coherence and QSD.

The relationship between quantifying quantum correlation and quantum state discrimination task is first revealed by Spehner and Orszag  \cite{spehner2013A,spehner2013B}. In coherence theory, the quantification of coherence has a closer link to QSD task which is described in the following theorem.

\begin{thm}\label{thm1}
Let $\rho$ be a state of the quantum system with d-dimensional Hilbert space $\mathcal{H}$ and let $\{\ket{i}\}^d_{i=1}$ be a reference orthonormal basis. The geometric coherence of $\rho$ is equal to the minimum error probability to discriminate $\{\ket{\psi_i},\eta_i\}^d_{i=1}$ with von Neumann measuremnt,
\begin{align}
C_g(\rho)=P^{opt v.N}_E(\{\ket{\psi_i},\eta_i\}^d_{i=1}),\nonumber
\end{align}
 where $\eta_i=\bra{i}\rho\ket{i}$ and $\ket{\psi_i}=\eta^{-1/2}_i\sqrt{\rho}\ket{i}$. Moreover, if the set of pure states $\{\ket{\psi_i}\}_i$ is linearly dependent, the geometric coherence provides an upper bound for the minimum error probability to discriminate $\{\ket{\psi_i},\eta_i\}^d_{i=1}$,
\begin{align}\label{eq11}
C_g(\rho)\ge P^{opt}_E(\{\ket{\psi_i},\eta_i\}^d_{i=1}).
\end{align}
If $\{\ket{\psi_i}\}_i$ is linearly independent, $C_g(\rho)$ is exactly equal to the minimum error probability, that is,
\begin{align}\label{eq3}
C_g(\rho)=P^{opt}_E(\{\ket{\psi_i},\eta_i\}^d_{i=1}).
\end{align}
\end{thm}

\begin{proof}

Firstly, we evaluate the geometric coherence $C_g(\rho)$.

Assuming $\sigma=\sum_i\mu_i\ket{i}\bra{i}$ is an arbitrary incoherent state, one has
\begin{align}\label{eq6}
\sqrt{F(\rho)}&=\mathop{\max}_{\sigma\in\mathcal{I}}||\sqrt{\rho}\sqrt{\sigma}||_1\nonumber\\
&=\mathop{\max}_{\sigma\in\mathcal{I}}\mathop{\max}_U|\mathrm{Tr}(U\sqrt{\rho}\sqrt{\sigma})|\nonumber\\
&=\mathop{\max}_{\{\mu_i\}}\mathop{\max}_U|\sum_i\sqrt{\mu_i}\bra{i}U\sqrt{\rho}\ket{i}|\nonumber\\
&=\mathop{\max}_{\{\mu_i\}}\mathop{\max}_{\ket{f_i}}\sum_i|\sqrt{\mu_i}\bra{f_i}\sqrt{\rho}\ket{i}|,
\end{align}
where $\ket{f_i}=U\ket{i},i=1,...,d$ and the maximal being over all incoherent states and all orthogonal basis on $\mathcal{H}$. The last $"="$ in (\ref{eq6}) is because one can choose the phase factors of $\ket{f_i}$ in such a way that $\bra{i}U\sqrt{\rho}\ket{i}\ge0$, thus the two expressions are in fact equal.

Thanks to Cauchy-Schwartz inequality, the maximum over the probability $\mu_i$ is reached for
\begin{align}\label{eq1}
\mu_i=\frac{|\bra{f_i}\sqrt{\rho}\ket{i}|^2}{\sum_i|\bra{f_i}\sqrt{\rho}\ket{i}|^2},
\end{align}
and
\begin{align}\label{eq7}
F(\rho)&=\mathop{\mathrm{max}}_{\ket{f_i}}\sum^d_{i=1}|\bra{f_i}\sqrt{\rho}\ket{i}|^2\nonumber\\
&=\mathop{\mathrm{max}}_{\{\Pi_i\}}\sum^d_{i=1}\mathrm{Tr}(\Pi_i\sqrt{\rho}\ket{i}\bra{i}\sqrt{\rho}),
\end{align}
where $\Pi_i=\ket{f_i}\bra{f_i}$. Therefore, $\{\Pi_i\}^d_{i=1}$ is a von Neumann measurement in $\mathcal{H}$.

We thus obtain
\begin{align}
F(\rho)=P^{opt v.N}_S(\{\ket{\psi_i},\eta_i\}^d_{i=1})\equiv\max_{\{\Pi_i\}}\sum_i\mathrm{Tr}(\Pi_i\ket{\psi_i}\bra{\psi_i}),\nonumber
\end{align}
where $\eta_i=\bra{i}\rho\ket{i}$, $\ket{\psi_i}=\eta^{-1/2}_i\sqrt{\rho}\ket{i}$ and $P^{opt v.N}_S$ represents the maximum successful probability over all von Neumann measurements in $\mathcal{H}$.

Secondly, we consider the ensemble $\{\ket{\psi_i},\eta_i\}^d_{i=1}$.

 If $\eta_i\ne0$, for $i=1,...,d$, which means that the ensemble contains $d$ states. There are two cases,
  on one hand, $\{\ket{\psi_i}\}^d_{i=1}$ linearly independent, then the optimal measurement is von Neumann measurement \cite{kennedy1973}, that is,
  \begin{align}
  P^{opt}_S(\{\ket{\psi_i},\eta_i\}^d_{i=1})=P^{opt v.N}_S(\{\ket{\psi_i},\eta_i\}^d_{i=1})=F(\rho).\nonumber
  \end{align}

  On the other hand, $\{\ket{\psi_i}\}^d_{i=1}$ is linearly dependent, which means that the optimal measurement is a general POVM. As a result, one has
  \begin{align}
  P^{opt}_S(\{\ket{\psi_i},\eta_i\}^d_{i=1})\ge P^{opt v.N}_S(\{\ket{\psi_i},\eta_i\}^d_{i=1})=F(\rho).\nonumber
  \end{align}

 If $\eta_i=0$ for some $i=i_1,i_2,...,i_s$, then
 the ensemble  $\{\ket{\psi_i},\eta_i\}$ reduces to $\{\ket{\psi_{i^{\prime}}},\eta_{i^{\prime}}\}^{d-s}_{i^{\prime}=1}$ with $\{\eta_{i^{\prime}}\}^{d-s}_{i^{\prime}=1}$ is the non-increasing array of non-zero $\eta_i$s. In fact, since $\eta_i=\bra{i}\rho\ket{i}=||\sqrt{\rho}\ket{i}||^2$, $\eta_i=0$ means that $\sqrt{\rho}\ket{i}$ is a $0$ vector. The same as above, there is two cases, linearly independent or not.
 In the first case, as
 \begin{align}
 F(\rho)&=\max_{\ket{f_i}}\sum^d_{i=1}|\bra{f_i}\sqrt{\rho}\ket{i}|^2\nonumber\\
 &=\max_{\ket{f_{i^{\prime}}}}\sum^{d-s}_{i^{\prime}=1}|\bra{f_{i^{\prime}}}\sqrt{\rho}\ket{i^{\prime}}|^2\nonumber\\
 &=P^{opt v.N}_S(\{\ket{\psi_{i^{\prime}}},\eta_{i^{\prime}}\}^{d-s}_{i^{\prime}=1}),\nonumber
 \end{align}
 one still has
 \begin{align}
 F(\rho)=P^{opt}_S(\{\ket{\psi_{i^{\prime}}},\eta_{i^{\prime}}\}^{d-s}_{i^{\prime}=1}).\nonumber
 \end{align}
In another case, $F(\rho)$ is also a lower bound for $P^{opt}_S$,
\begin{align}
F(\rho)\le P^{opt}_S(\{\ket{\psi_{i^{\prime}}},\eta_{i^{\prime}}\}^{d-s}_{i^{\prime}=1}). \nonumber
\end{align}

In conclusion, for any $\rho$, $C_g(\rho)$ provides an upper bound for $P^{opt}_E(\{\ket{\psi_i},\eta_i\})$, namely
\begin{align}
C_g(\rho)\ge P^{opt}_E(\{\ket{\psi_i},\eta_i\}).
\end{align}

 Especially, if $\{\ket{\psi_i},\eta_i\}$ is linearly independent, $C_g(\rho)$ is exactly the minimum error probability of QSD,
\begin{align}
C_g(\rho)=P^{opt}_E(\{\ket{\psi_i},\eta_i\}).
\end{align}

\end{proof}

Based on (\ref{eq1}) and (\ref{eq7}), the CIS of $\rho$ is given by
\begin{align}
\sigma_{\rho}=\frac{1}{F(\rho)}\sum_i\bra{i}\sqrt{\rho}\Pi^{opt}_i\sqrt{\rho}\ket{i}\ket{i}\bra{i},\nonumber
\end{align}
where $\{\Pi^{opt}_i\}^d_{i=1}$ is the optimal von Neumann measurement.

\begin{cor}
If $\rho>0$, the geometric coherence is equal to the minimum error probability to discriminate $\{\ket{\psi_i},\eta_i\}^d_{i=1}$, that is,
\begin{align}\label{eq23}
C_g(\rho)=P^{opt}_E(\{\ket{\psi_i},\eta_i\}^d_{i=1}).
\end{align}
\end{cor}
\begin{proof}
If $\rho>0$, then the collection of $\{\ket{\psi_i}\}^d_{i=1}$ is linearly independent. In fact, if it is not the case, then there exist a non-zero vector $(x_1,x_2,...,x_d)^t$, such that $\sum_ix_i\sqrt{\rho}\ket{i}=0$. That is, $\sqrt{\rho}(\sum_ix_i\ket{i})=0$ for a non-zero vector $\sum_ix_i\ket{i}$, which is conflict with the condition $\rho>0$. With theorem \ref{thm1}, we obtain \eqref{eq23}.

\end{proof}

\section{discriminate quantum states with geometric coherence}

We link the geometric coherence to the quantum state discrimination in last section. How about the opposite situation? That is, given a pure QSD ensemble $\{\ket{\psi_i},\eta_i\}^d_{i=1}$, is there exist a quantum state whose geometric coherence provides an upper bound for the minimum error probability of discrimination?

Let us consider a pure state discrimination $\{\ket{\psi_i},\eta_i\}^d_{i=1}$. Denoting a matrix $\rho$ with $\rho_{ij}=\sqrt{\eta_i\eta_j}\langle\psi_i|\psi_j\rangle$, $1\le i,j\le d$, that is,
 \begin{align}\label{eq10}
 \rho=\begin{pmatrix}
\eta_1&\sqrt{\eta_1\eta_2}\langle\psi_1|\psi_2\rangle&...& \sqrt{\eta_1\eta_d}\langle\psi_1|\psi_d\rangle\\
\sqrt{\eta_2\eta_1}\langle\psi_2|\psi_1\rangle&\eta_2&...&\sqrt{\eta_2\eta_d}\langle\psi_2|\psi_d\rangle\\
.&.&...&.\\.&.&...&.\\
\sqrt{\eta_d\eta_1}\langle\psi_d|\psi_1\rangle&\sqrt{\eta_d\eta_2}\langle\psi_d|\psi_2\rangle&...&\eta_d
\end{pmatrix}.
\end{align}

\begin{prop}
 The matrix $\rho$ is a density matrix.
\end{prop}
\begin{proof}
Denoting a matrix $\psi$ with each entry $\psi_{ij}=\sqrt{\eta_j}(\psi_j)_i$ with $(\psi_j)_i$ the $i$-th entry of $\ket{\psi_j}$, then the matrix
\begin{align}
\rho=\psi^{\dagger}\psi\ge0\nonumber
\end{align}
is positive semidefinite.
As $\sum_i\rho_{ii}=\sum_i\eta_i=1$, we conclude that $\rho$ is a density matrix.
\end{proof}

Therefore, we call the state \eqref{eq10} the QSD-state of $\{\ket{\psi_i},\eta_i\}^d_{i=1}$.

Based on Theorem \ref{thm1}, the corresponding QSD ensemble of $\rho$ is $\{\ket{\phi_i},p_i\}^d_{i=1}$, where
\begin{align}
\ket{\phi_i}=p^{-1/2}_i\sqrt{\rho}\ket{i},p_i=\rho_{ii}.\nonumber
\end{align}

Therefore, one has
\begin{align}
\langle\phi_i\ket{\phi_j}=\frac{\rho_{ij}}{\sqrt{p_ip_j}}=\langle\psi_i\ket{\psi_j},1\le i,j\le d\nonumber.
\end{align}

About the discrimination task $\{\ket{\psi_i},\eta_i\}^d_{i=1}$ and $\{\ket{\phi_i},\eta_i\}^d_{i=1}$, we have the following result.

\begin{prop}\label{prop1}
If $\langle \psi_i\ket{\psi_j}=\langle \phi_i\ket{\phi_j}$( or $\overline{\langle \phi_i\ket{\phi_j}}$), $1\le i\ne j\le d$, then $P^{opt }_S(\{\ket{\psi_i},\eta_i\}^d_{i=1})=P^{opt }_S(\{\ket{\phi_i},\eta_i\}^d_{i=1})$.
\end{prop}

\begin{proof}
See in appendix \ref{app1}.
\end{proof}

For $U\in U(d)$, $\{U\ket{\psi_i},\eta_i\}^d_{i=1}$ have the same QSD-state. Actually, these quantities $\{\langle\psi_i|\psi_j\rangle,\eta_i,1\le i,j\le d\}$ contain all the information about the discrimination of $\{\ket{\psi_i},\eta_i\}^d_{i=1}$.

In conclusion, we have the following result.
\begin{thm}\label{thm2}
Let $\mathcal{H}$ be a $d$-dimensional Hilbert space and $\{\ket{i}\}^d_{i=1}$ be the computable basis, that is, $\ket{i}=(0,...,0,1,0,...,0)^t$, the $i$-th entry is $1$ for $i=1,...,d$. For $\ket{\psi_i}\in\mathcal{H},i=1,...,d$, the minimal error probability to discriminate the collection of pure states $\{\ket{\psi_i},\eta_i\}^d_{i=1}$ is upper bounded by the geometric coherence of the corresponding QSD-state $\rho$, that is,
\begin{align}
P^{opt}_E(\{\ket{\psi_i},\eta_i\}^d_{i=1})\le C_g(\rho),
\end{align}
where the incoherent pure states are $\{\ket{i}\}^d_{i=1}$. In particular, if the set of pure states is linearly independent, the bound is reached. That is,
\begin{align}\label{eq19}
P^{opt}_E(\{\ket{\psi_i},\eta_i\}^d_{i=1})= C_g(\rho).
\end{align}

\end{thm}
\begin{proof}

Based on the proof of Theorem \ref{thm1}, we have that
\begin{align}
C_g(\rho)=P^{opt v.N}_E(\{\ket{\phi_i},p_i\}^d_{i=1}),\nonumber
\end{align}
where $\ket{\phi_i}=p^{-1/2}_i\sqrt{\rho}\ket{i}$ and $p_i=\rho_{ii}=\eta_i$. Since
\begin{align}
\langle\phi_i|\phi_j\rangle=\frac{1}{\sqrt{\eta_i\eta_j}}\rho_{ij}=\langle\psi_i|\psi_j\rangle,\forall i,j,\nonumber
\end{align}
then,
\begin{align}
P^{opt}_E(\{\ket{\psi_i}, \eta_i\}^d_{i=1})=P^{opt}_E(\{\ket{\phi_i}, \eta_i\}^d_{i=1}).\nonumber
\end{align}
 Therefore, we have
 \begin{align}
 C_g(\rho)\ge P^{opt}_E(\{\ket{\psi_i}, \eta_i\}^d_{i=1}),\nonumber
 \end{align}
 for linearly dependent $\{\ket{\psi_i}\}^d_{i=1}$ and
 \begin{align}
 C_g(\rho)=P^{opt}_E(\{\ket{\psi_i}, \eta_i\}^d_{i=1}),\nonumber
 \end{align}
 for linearly independent $\{\ket{\psi_i}\}^d_{i=1}$.
\end{proof}

\begin{rem}  With (\ref{eq19}), we can see that more quantumness the QSD-state contains, more difficult the QSD task is. Moreover, the QSD-state is incoherent, that is, classical, if and only if there exist a quantum measurement to discriminate these states perfectly. Theorem \ref{thm2} indicates that non-orthogonality of states lies at very heart of quantum mechanics from the perspective of coherence theory.
\end{rem}

\section{duality relationbetween geometric coherence and path distinguishability}

\citeauthor{Bera2015} reveal the complementarity of coherence and path distinguishability in the case of Yang's $n$-slit experiment. Yang's experiment can be explained by wave-particle duality. In \cite{Bera2015}, the authors quantify wave and particle nature with $l_1$- norm of coherence and unambiguous state discrimination, respectively. With theorem \ref{thm1}, we now establish a complementarity relationship of geometric coherence and path distinguishability as follows.

Similarly, let us first consider the case of $d$-slit quantum interference with pure quantons. Denoting $\ket{i}$ as the possible state of the quanton when it takes the $i$th slit or $i$th path, then the state of quanton can be represented with $d$ basis states $\{\ket{1},...,\ket{d}\}$ as
\begin{align}
\ket{\Psi}=c_1\ket{1}+...+c_d\ket{d},
\end{align}
where $\ket{i}$ represents the $i$th slit and $c_i$ is the amplitude of taking the $i$th slit. To know which path the quanton passes, one needs to perform a quantum measurement. According to quantum measurement theory, the quanton will interact with a detector state and get entangled as
\begin{align}
U(\ket{\Psi}\ket{0_d})=\sum_ic_i\ket{i}\ket{d_i}
\end{align}
with $\{\ket{d_i}\}$ are normalized and not necessarily orthogonal.

On one hand, to know the coherence of quanton, one considers the reduced density matrix of the quanton after tracing out the detector states,
\begin{align}
\rho_s=\sum^d_{i,j=1}c_i\bar{c}_j\langle d_j\ket{d_i}\ket{i}\bra{j}.
\end{align}

With Theorem \ref{thm1}, one has
\begin{align}
C_g(\rho_s)=1-P^{opt v.N}_S(\{\ket{\psi_i},\eta_i\}^d_{i=1}),
\end{align}
where $\eta_i=|c_i|^2,\ket{\psi_i}=\exp(\sqrt{-1}\theta_i)\eta^{-1/2}_i\sqrt{\rho_s}\ket{i}$ and $\theta_i$ is the argument of $c_i$.

On the other hand, if one wants to know which path it takes, he needs to discriminate the detector states $\{\ket{d_i},|c_i|^2\}^d_{i=1}$. In other words, the path distinguishability is actually equivalent to the discrimination of detector states.

With Proposition \ref{prop1} and the fact $\langle \psi_i\ket{\psi_j}=\langle d_j\ket{d_i}$, one has that
\begin{align}
P^{opt}_S(\{\ket{\psi_i},|c_i|^2\}^d_{i=1})=P^{opt }_S(\{\ket{d_i},|c_i|^2\}^d_{i=1}).\nonumber
\end{align}

Defining the optimal successful probability to discriminate the detector states as path distinguishability,
\begin{align}
D_q:=P^{opt}_S(\{\ket{d_i},\rho_{ii}\}^d_{i=1}),
\end{align}
and the geometric coherence as coherence,
\begin{align}
C_g:=C_g(\rho_s).
\end{align}


Since the von Neumann measurement is the optimal if and only if the detector states is linearly independent \cite{kennedy1973}, then for linearly independent $\{\ket{d_i}\}$, we have the following complementarity relation,
\begin{align}\label{eq21}
C_g+D_q=1.
\end{align}

Above complementarity relation can be generalized to the situation that the quanton state is a mixed state, say $\rho=\sum_{i,j}\rho_{ij}\ket{i}\bra{j}$. This is the case that the quantum system is exposed to the environment. The combined system of the quanton and the path detector after the measurement interaction can be written as
\begin{align}
\rho_{sd}=\sum_{i,j}\rho_{ij}\ket{i}\bra{j}\otimes\ket{d_i}\bra{d_j},
\end{align}
and the reduced density matrix of the quanton after tracing out the detector states,
\begin{align}\label{eq24}
\rho_s=\sum^d_{i,j=1}\rho_{ij}\langle d_j\ket{d_i}\ket{i}\bra{j}.
\end{align}

Therefore, on one hand, the wave nature of the quanton can be characterize by the coherence of $\rho_s$, namely $C_g(\rho_s)$. As every principle $2\times2$ submatrix of \eqref{eq24} is positive semidefinite \cite[p.434]{Horn:2012:MA:2422911}, we have
\begin{align}
\sqrt{\rho_{ii}\rho_{jj}}-|\rho_{ij}|\ge0, 1\le i,j\le d.
\end{align}
Assuming the corresponding ensemble to $\rho_s$ is $\{\ket{\psi_i},\rho_{ii}\}$, then
\begin{align*}
|\langle\psi_i\ket{\psi_j}|=\frac{|\bra{i}\rho_s\ket{j}|}{\sqrt{\rho_{ii}\rho_{jj}}}=\frac{|\rho_{ij}|}{\sqrt{\rho_{ii}\rho_{jj}}}|\langle d_i\ket{d_j}|\le|\langle d_i\ket{d_j}|,
\end{align*}
for each $i,j$. In other words, each two states of $\{\ket{d_i},\rho_{ii}\}^d_{i=1}$ is more difficult to distinguished than the corresponding of $\{\ket{\psi_i},\rho_{ii}\}^d_{i=1}$, then
\begin{align*}
P^{opt}_S(\{\ket{\psi_i},\rho_{ii}\}^d_{i=1})\ge P^{opt}_S(\{\ket{d_i},\rho_{ii}\}^d_{i=1}).
\end{align*}

On the other hand, we quantify the particle nature with path distinguishability, that is, the optimal successful probability to discriminate $\{\ket{d_i},\rho_{ii}\}^d_{i=1}$. In conclusion, for linearly independent ensemble $\{\ket{d_i},\rho_{ii}\}^d_{i=1}$, the following complementarity relation holds,
\begin{align*}
C_g+D_q\le1.
\end{align*}

Besides, we can derive a relationship between geometric coherence and $l_1$-norm of coherence. In \cite{Bera2015}, the authors choose unambiguous quantum state discrimination as the optimal strategy to distinguish detector states and establish the complementarity relation as
\begin{align}\label{eq22}
C+D_Q\le1,
\end{align}
where $C:=\frac{C_{l_1}(\rho_s)}{d-1}$ with $C_{l_1}(\rho):=\sum_{i\ne j}|\bra{i}\rho\ket{j}|$ and $D_Q��=P^{opt,u}_S(\{\ket{d_i},\rho_{ii}\}^d_{i=1})$ is the optimal successful probability to unambiguously discriminate the detector states. If the quanton is pure, \eqref{eq22} is a equality.

Comparing \eqref{eq21} and \eqref{eq22}, we have the following theorem.

\begin{thm}\label{thm3}
Let $\rho$ be a density matrix in a $d$-dimensional Hilbert space and $\{\ket{\psi_i},\eta_i\}^d_{i=1}$ be the corresponding discrimination task. If $\{\ket{\psi_i},\eta_i\}^d_{i=1}$ is linearly independent, then
\begin{align}
\frac{C_{l_1}(\rho)}{d-1}\ge C_g(\rho).
\end{align}
\end{thm}
\begin{proof}
For any $\rho$, it can be represented as
\begin{align}
\rho&=\sum_{i,j}\sqrt{\rho_{ii}}\sqrt{\rho_{jj}}\frac{\bra{i}\sqrt{\rho}}{\sqrt{\rho_{ii}}}\frac{\sqrt{\rho}\ket{j}}{\sqrt{\rho_{jj}}}\ket{i}\bra{j}\nonumber\\
&=\sum_{i,j}c_ic_j\langle d_j\ket{d_i}\ket{i}\bra{j},\nonumber
\end{align}
with $c_i=\sqrt{\rho_{ii}}$ and $\ket{d_i}=\frac{\sqrt{\rho}\ket{i}}{\sqrt{\rho_{ii}}},1\le i\le d.$ Since $\{\ket{d_i}\}$ is linearly independent, one has
\begin{align*}
C_g(\rho)+P^{opt}_S(\{\ket{d_i},\rho_{ii}\}^d_{i=1})=1,
\end{align*}
and
\begin{align*}
\frac{C_{l_1}(\rho)}{d-1}+P^{opt,u}_S(\{\ket{d_i},\rho_{ii}\}^d_{i=1})=1.
\end{align*}

On the other hand, as the successful probability with unambiguous QSD will never exceed the maximal successful probability with optimal measurement, that is,
\begin{align}
P^{opt}_S(\{\ket{d_i},\rho_{ii}\}^d_{i=1})\ge P^{opt,u}_S(\{\ket{d_i},\rho_{ii}\}^d_{i=1}),\nonumber
\end{align}
one has
\begin{align}\label{eq2}
C_g(\rho)\le \frac{C_{l_1}(\rho)}{d-1}.
\end{align}
\end{proof}

\section{evaluate geometric coherence with QSD}

Based on the equivalence of geometric coherence and quantum state discrimination, we will give an upper bound for the former and calculate geometric coherence for a set of states in this section.

\subsection{a tighter upper bound for geometric coherence}

The analytical expression of geometric coherence for any single-qubit state $\rho$ is given in \cite{Streltsov2015B}. However, the computation of $C_g$ is formidably difficult for even qudit state. With sub-fidelity introduced as a lower bound for quantum fidelity \cite{Miszczak2009}, the authors give an upper bound for geometric coherence as follows.
\begin{thm}\cite{zhanghj2017}
Let $\mathcal{H}$ be a finite dimensional Hilbert space and $\rho\in\mathcal{E}(\mathcal{H})$ be a quantum state, then
\begin{align}
C_g(\rho)\le\min\{l_1,l_2\},
\end{align}
where $l_1=1-\max_i\{\rho_{ii}\}$ and $l_2=1-\sum_ib^2_{ii}$ with $b_{ij}$ is the $(i,j)$-th entry of $\sqrt{\rho}$.
\end{thm}

 Based on Theorem \ref{thm2}, we can derive a tighter upper bound for geometric coherence.

 Above all, rearrange $\{\ket{\psi_i},\eta_i\}^d_{i=1}$, the corresponding discrimination task of $\rho$, into $\{\ket{\varphi_i},\xi_i\}^d_{i=1}$ with $\xi_1\ge...\ge\xi_{d}$. Let $d^{\prime}$ be the dimension of the space spanned by $\{\ket{\psi_1},...,\ket{\psi_d}\})$, we can choose an independent ensemble $\{\ket{\varphi_i},\xi_i\}^{d^{\prime}}_{i=1}$ as follows. First step, we choose $\ket{\varphi_1}$ such that $\xi_1=\max\{\eta_1,...,\eta_d\}$. In step $n$ ($2\le n\le d^{\prime}$), choose $\ket{\varphi_n}$ such that $\{\ket{\varphi_1},...,\ket{\varphi_{n-1}},\ket{\varphi_n}\}$ is linearly independent and $\xi_n=\max\{\eta_1,...,\eta_d\}/\{\xi_1,...,\xi_{n-1}\}$. In this way, we obtain a linearly independent ensemble $\{\ket{\varphi_i},\xi_i\}^{d^{\prime}}_{i=1}$.

Then, we construct a measurement to discriminate $\{\ket{\varphi_i},\xi_i\}^d_{i=1}$ with the Gram-Schmidt orthogonalization (GSO) of $\{\ket{\phi_i}\}^{d^{\prime}}_{i=1}$. First of all, we make the GSO for $\{\ket{\phi_i}\}^{d^{\prime}}_{i=1}$ as
\begin{align}
\ket{\phi_1}&=\ket{\varphi_1},\nonumber\\
\ket{\phi_2}&=\frac{P_2\ket{\varphi_2}}{\sqrt{\bra{\varphi_2}P_2\ket{\varphi_2}}} \nonumber\\
&......\nonumber\\
\ket{\phi_{d^{\prime}}}&=\frac{P_{d^{\prime}}\ket{\varphi_{d^{\prime}}}}{\sqrt{\bra{\varphi_{d^{\prime}}}P_{d^{\prime}}\ket{\varphi_{d^{\prime}}}}},\nonumber
\end{align}
where $P_n=I_n-\sum^{n-1}_{i=1}\ket{\phi_i}\bra{\phi_i}$ is the projection to the normal direction of spanned by $\{\ket{\varphi_1},...,\ket{\varphi_{n-1}}\}$ for $n=2,..,d$.

Secondly, we extend $\{\ket{\phi_1},...,\ket{\phi_{d^{\prime}}}\}$ to $\{\ket{\phi_1},...,\ket{\phi_d}\}$ through adding $(d-d^{\prime})$ unit vectors such that $\sum^d_i\ket{\phi_i}\bra{\phi_i}=I_d$.
As a result, $\{\ket{\phi_1}\bra{\phi_i}\}^d_{i=1}$ is a von Neumann measurement and the corresponding successful probability to discriminate $\{\ket{\psi_i},\eta_i\}^d_{i=1}$ is equal to
\begin{align}\label{eq16}
P^{gso}_S(\{\ket{\varphi_i},\xi_i\}^d_{i=1})&:=\sum^d_{i=1}\xi_i|\langle{\varphi_i}\ket{\phi_i}|^2\nonumber\\
&=\sum^{d^{\prime}}_{i=1}\xi_i(1-\sum^{i-1}_{j=1}|\langle{\varphi_i}\ket{\phi_j}|^2)\nonumber\\
&=1-\sum^{d^{\prime}}_{i=1}\sum^{i-1}_{j=1}\xi_i|\langle{\varphi_i}\ket{\phi_j}|^2.
\end{align}

Obviously, for ${d^{\prime}}\ge2$, we have
\begin{align}
P^{gso}_E(\{\ket{\psi_i},\xi_i\}^d_{i=1})&=1-(\sum^{d^{\prime}}_{i=1}\xi_i(1 -\sum^{i-1}_{j=1}|\langle{\psi_i}\ket{\phi_j}|^2))\nonumber\\
&< 1-\xi_1=1-\max_i\rho_{ii}=l_1.
\end{align}

With Theorem \ref{thm1}, one has
\begin{align}
C_g(\rho)= P^{opt v.N}_E(\{\ket{\psi_i},\eta_i\}^d_{i=1})\le P^{gso}_E(\{\ket{\psi_i},\eta_i\}^d_{i=1}) \nonumber
\end{align}

Therefore, for $d^{\prime}\ge2$, we obtain a tighter bound than $l_1$ as
\begin{align}\label{eq17}
P^{gso}_E(\{\ket{\psi_i},\eta_i\}^d_{i=1})=\sum^{d^{\prime}}_{i=1}\sum^{i-1}_{j=1}\xi_i|\langle{\varphi_i}\ket{\phi_j}|^2.
\end{align}

In conclusion, we have the following theorem.
\begin{thm}
Let $\mathcal{H}$ be a finite dimensional Hilbert space and $\rho\in\mathcal{E}(\mathcal{H})$ be a quantum state, then
\begin{align}
C_g(\rho)\le\min\{l_2,l_3\},
\end{align}
where $l_3$ denote the upper bound in \eqref{eq17} and $l_2=1-\sum_ib^2_{ii}$ with $b_{ij}$ is the $(i,j)$-th entry of $\sqrt{\rho}$. When the Gram-Schmidt orthogonalization is an optimal measurement, the upper bound $l_3$ is tight.
\end{thm}

\begin{rem}
If $\rho>0$, the upper bound for geometric coherence can be improved as
\begin{align}
C_g(\rho)\le\min\{l_2,l_3,l_4\},
\end{align}
where $l_4=\frac{C_{l_1}(\rho)}{d-1}$. Moreover, the upper bound $l_4$ is tight when the corresponding measurement to unambiguously discriminate $\{\ket{\psi_i},\eta_i\}^d_{i=1}$ is also a optimal measurement.
\end{rem}

\subsection{geometric coherence for generalized X-state}

We can give the analytical formula of geometric coherence for a class of quantum states on any finite dimensional Hilbert space with the result from QSD.

First of all, we consider the single-qubit state. The simplest example of ambiguous discrimination is the case of two pure states $\ket{\psi_1}$ and $\ket{\psi_2}$. Then the optimal success probability is easy to determine as \cite{Helstrom1976}
\begin{align}\label{eq4}
P^{opt}_S(\{\rho_i,\eta_i\}^2_{i=1})=\frac{1}{2}(1+\sqrt{1-4\eta_1\eta_2|\langle\psi_1|\psi_2\rangle|^2}).
\end{align}

With (\ref{eq4}) and Theorem \ref{thm1}, the geometric coherence for one qubit state can be evaluated without any thinking. Since any a coherent single-qubit state is invertible, one has,
\begin{align}\label{eq5}
C_g(\rho)&=\frac{1}{2}(1-\sqrt{1-4\rho_{11}\rho_{22}|(\rho_{11}\rho_{22})^{-1/2}\bra{1}\rho\ket{2}|^2}),\nonumber\\
&=\frac{1}{2}(1-\sqrt{1-4|\rho_{12}|^2}).
\end{align}

Secondly, we generalize this result to X-state in higher dimensional space. Any quantum state $\rho$ is called X-state if it can be represented as an X-type matrix in a fixed orthogonal basis $\{\ket{i}\}^d_{i=1}$ as
\begin{align}\label{eq8}
\rho=\begin{pmatrix}
\rho_{11} &0&.&.&0&\rho_{1d}\\0&\rho_{22}&.&.&\rho_{2,d-1}&0\\.&.&.&.&.&.\\
.&.&.&.&.&.\\0&\rho_{d-1,2}&.&.&\rho_{d-1,d-1}&0\\ \rho_{d1} &0&.&.&0&\rho_{dd}
\end{pmatrix}.
\end{align}

In two-qubit case, X-states are a significant class of states including Bell-diagonal states which play an important role in the quantification and dynamics of entanglement, quantum correlations and coherence \cite{Vedral1997,luo2008A,Dakic2010,M.Ali2010A,Rau2009,Bromley2015}. The geometric coherence of (\ref{eq8}) is equal to the minimal error probability to discriminate $\{\ket{\psi_i},\rho_{ii}\}^d_{i=1}$, where $\ket{\psi_i}=\rho^{-1/2}_{ii}\sqrt{\rho}\ket{i}$. In fact, as
\begin{align}
\langle\psi_i|\psi_j\rangle=\frac{1}{\sqrt{\rho_{ii}\rho_{jj}}}\rho_{ij}\nonumber,
\end{align}
it is easy to see that, each $\ket{\psi_i}$ is orthogonal to others $\ket{\psi_j}$ except $\ket{\psi_{d+1-i}}$. If $d>2$ is even (odd case is similar) , the collection of $\{\ket{\psi_i}\}^d_{i=1}$ can be divided into $d/2$ orthogonal groups, $\{\ket{\psi_1},\ket{\psi_n}\},...,\{\ket{\psi_{d/2-1}},\ket{\psi_{d/2}}\}$ in which each two states in a group are non-orthogonal and the states in different groups are orthogonal. In other words, we just need to discriminate two non-orthogonal states in each group one by one. Therefore, with (\ref{eq4}), the geometric coherence of X-state $\rho$ is easy to obtain,
\begin{align}
C_g(\rho)=1-\frac{1}{2}\sum^{d/2}_i(\rho_{ii}+\rho_{d+1-i,d+1-i})(1+\sqrt{1-|\rho_{i,d+1-i}|^2}).\nonumber
\end{align}

In the end, we consider the generalized X-state, that is these states can be converted to X-states by permutation transformations. In other words, generalized X-states are these states that have at most only one non-zero entry in each column and each row in non-diagonal part. The geometric coherence of generalized X-states can be also evaluated similarly,
\begin{align}\label{eq18}
C_g(\rho)=1-\frac{1}{4}\sum^{d}_i(\rho_{ii}+\rho_{\bar{i},\bar{i}})(1+\sqrt{1-|\rho_{i,\bar{i}}|^2}),
\end{align}
where $\rho_{i,\bar{i}}$ is the only possible non-zero entry in each row except the diagonal entry.

Besides, geometric coherence of generalized X-states can be obtained from the geometric coherence of X-state with the incoherent unitary-invariant of $C_g$.

\section{discriminate states with geometric coherence}

On the contrast, we consider the problem of quantum state discrimination with the help of coherence theory. There are two examples, QSD with multiple copies and discriminate two pure states with geometric coherence.

\subsection{QSD with multiple copies}

Even though it is impossible to distinguish non-orthogonal states perfectly, the error of possibility can be decreased if there are more copies of states.

Let us consider the QSD task of n copies of pure states, that is $\{\ket{\psi_i}^{\otimes n},\eta_i\}^d_{i=1}$. Based on theorem \ref{thm2}, the $i,j$-th entry of corresponding QSD-state is
\begin{align}
\rho^{(n)}_{ij}=\sqrt{\eta_i\eta_j}\langle\psi_i\ket{\psi_j}^n,\forall 1\le i,j\le d.\nonumber
\end{align}

For big $n$, it is easy to check that $\{\ket{\psi_1}^{\otimes n},...,\ket{\psi_d}^{\otimes n}\}$ is linearly independent, then $\rho^{(n)}$ is invertible. Thanks to Theorem \ref{thm3}, one has that
\begin{align}
C_g(\rho^{(n)})\le\frac{C_{l_1}(\rho^{(n)})}{d-1}.
\end{align}

As $n\rightarrow\infty$, each $\rho^{(n)}_{ij}\rightarrow0$ for $i\ne j$ and so is $C_{l_1}(\rho^{(n)})$. That is to say, if we have enough copies, we can discriminate $\{\ket{\psi_1},...,\ket{\psi_d}\}$ almost perfectly.

In fact, since $\{\ket{\psi_1}^{\otimes n},...,\ket{\psi_1}^{\otimes n}\}$ is linearly independent, the corresponding Gram-Schmidt orthogonalization can be a suitable measurement.

For simplicity, let us we consider $d=3$ case. With \eqref{eq16}, one has
\begin{align}
&P^{gso}_S(\{\ket{\psi_i}^{\otimes n},\eta_i\}^3_{i=1})=\xi_1+\eta_2(1-|\langle\psi_1\ket{\psi_2}|^{2n})\nonumber\\
+&\xi_3(1-|\langle\psi_1\ket{\psi_3}|^{2n}-\frac{|\langle\psi_3\ket{\psi_2}^n-\langle\psi_3\ket{\psi_1}^n\langle\psi_1\ket{\psi_2}^n|^2}{1-|\langle\psi_1\ket{\psi_2}|^{2n}})\nonumber\\
=&1-\frac{|\rho^{(n)}_{12}|^2}{\xi_1}-\frac{|\rho^{(n)}_{13}|^2}{\xi_1}-\frac{|\xi_1\rho^{(n)}_{32}-\rho^{(n)}_{31}\rho^{(n)}_{12}|^2}{\xi_1(\xi_1\xi_2-|\rho^{(n)}_{12}|^2)}.\nonumber
\end{align}

Let $n$ tends to $\infty$, the successful probability to discriminate n copies of ensembles tends to 1.

\subsection{discriminate quantum states with geometric coherence}\label{app3}

In the task of ambiguous quantum state discrimination, the task is to design the optimal measurement which maximized the success  probability to discriminate these states. In general, no effective method has been found to construct the optimal measurement to discriminate more than two states. In \cite{Helstrom1976}, Helstrom give the method to discriminate two pure states. We first briefly introduce Helstrom's method for two pure states and then give a new method with the equivalence between geometric and QSD. Moreover, we will show these two method in an example.

For two pure states QSD $\{\ket{\psi_i},\eta_i\}^2_{i=1}$. Assuming the measurement is $\{\ket{f_1},\ket{f_2}\}$, then the optimal successful probability is
\begin{align}\label{eq9}
P^{opt}_S(\{\ket{\psi_i},\eta_i\})=\sum^2_{i=1}\eta_i|\langle\psi_i\ket{f_i}|^2=\frac{1}{2}(1+\mathrm{Tr}|\Lambda|)
\end{align}
with $\Lambda=\eta_1\ket{\psi_1}\bra{\psi_1}-\eta_2\ket{\psi_2}\bra{\psi_2})$. The maximum achieved when $\ket{f_1}\bra{f_1}$ is the spectral projector $\Pi_1$ associated to the positive eigenvalue of Hermitian matrix $\Lambda$. \eqref{eq9} is called Helstrom formula.

 Furthermore, we introduce a new method to discriminate a set of linearly independent $\{\ket{\psi_i},\eta_i\}^d_{i=1}$  with Theorem \ref{thm2} and Lemma \ref{lem2}.

(1) Determining the ensemble $\{\ket{\psi^{\prime}_i},\eta_i\}^d_{i=1}$ for the corresponding QSD-state $\rho$, where $\ket{\psi^{\prime}_i}=\eta^{-1/2}_i\sqrt{\rho}\ket{i}, i=1,2,...,d$.

(2) Calculating the unitary matrix $U$ which transforms $\ket{\psi^{\prime}_i}$ to $\ket{\psi_i}$, namely, $\ket{\psi_i}=U\ket{\psi^{\prime}_i}$ for each $i$.

(3) Finding the optimal measurement $\{\ket{f^{\prime}_i}\}^d_{i=1}$, which optimally discriminate $\{\ket{\psi^{\prime}_i},\eta_i\}^d_{i=1}$, by solving the following equations,
\begin{align}
&\mu_i=\frac{|\bra{f^{\prime}_i}\sqrt{\rho}\ket{i}|^2}{F(\rho)},i=1,2,...,d,\nonumber\\
&\langle f^{\prime}_i\ket{f^{\prime}_j}=\delta_{ij},1\le i,j\le d.\nonumber
\end{align}

(4) Obtain the optimal measurement for $\{\ket{\psi_i},\eta_i\}^d_{i=1}$ as
\begin{align}
\ket{f_i}=U\ket{f^{\prime}_i},i=1,2,...,d.\nonumber
\end{align}

\begin{rem}
To evaluate geometric coherence, we have to calculate the eigenvalues of the state $\sqrt{\sigma}\rho\sqrt{\sigma}$ and maximize $\mathrm{Tr}(\sqrt{\sigma}\rho\sqrt{\sigma})$ over all incoherent states $\sigma$. The first part is actually the problem of finding the root of one element $n$ order equation. Since there is no root solution for the one element $n(\ge5)$ order equation, the $\mu_i$s may not be expressed explicitly for $n\ge5$ case. Even though, our method may shed a new light on the problem of distinguishing three linearly independent pure states which is an open problem that has lasted for many years.
\end{rem}

In Helstrom's method, the first step is determining the optimal measurement through finding the projector for positive eigenvalues and then obtain the optimal successful probability. In our result, we can get $P^{opt}_S$ directly without any knowledge of the optimal measurement, and the optimal measurement can be obtain through solving the equations about the closest incoherent states.

Now we compare our method with Helstrom's result in the following example. Let us consider the ensemble $\{\ket{\psi_i},\frac{1}{2}\}^2_{i=1}$ . Without of generality, one can choose an orthogonal basis $\{\ket{1},\ket{2}\}$ such that \cite{Barnett09},
\begin{align}
&\ket{\psi_1}=\cos\theta\ket{1}+\sin\theta\ket{2},\nonumber\\
&\ket{\psi_2}=\cos\theta\ket{1}-\sin\theta\ket{2}.\nonumber
\end{align}

One one hand, assuming the measurement is $\{\ket{f_1},\ket{f_2}\}$. Based on the Helstrom formula, the optimal successful probability is
\begin{align*}
P^{opt}_S(\{\ket{\psi_i},\frac{1}{2}\})=\frac{1}{2}(1+\mathrm{Tr}|\Lambda|)
\end{align*}
with $\Lambda=\frac{1}{2}(\ket{\psi_1}\bra{\psi_1}-\ket{\psi_2}\bra{\psi_2})$. Since
\begin{align*}
\Lambda=\begin{pmatrix}
0&\cos\theta\sin\theta\\
\cos\theta\sin\theta&0
\end{pmatrix},
\end{align*}
 the the optimal measurement is
\begin{align}\label{eq14}
 \ket{f_1}=\frac{1}{\sqrt{2}}(1,1)^t,\ket{f_2}=\frac{1}{\sqrt{2}}(1,-1)^t,
 \end{align}
and the optimal success probability is
\begin{align}\label{eq15}
P^{opt}_S(\{\ket{\psi_i},\frac{1}{2}\})&=\frac{1}{2}(1+\sqrt{1-\cos^22\theta})\nonumber\\
&=\frac{1}{2}(1+\sqrt{1-|\langle\psi_1\ket{\psi_2}|^2}).
\end{align}

One the other hand, let us determining the optimal successful probability and measurement with the above four steps. Since the corresponding QSD-state to $\{\ket{\psi_i},\frac{1}{2}\}^2_{i=1}$ is
\begin{align}
 \rho=\frac{1}{2}\begin{pmatrix}
1&\cos2\theta\\
\cos2\theta&1
\end{pmatrix},
\end{align}
 the maximal success probability to discriminate $\{\ket{\psi_i},\frac{1}{2}\}^2_{i=1}$ is easy to obtain as \cite{Streltsov2015B},
\begin{align}
F(\rho)=\frac{1}{2}(1+\sqrt{1-\cos^22\theta})=\frac{1}{2}(1+\sqrt{1-|\langle\psi_1\ket{\psi_2}|^2}),\nonumber
\end{align}
which is coincide with (\ref{eq15}). Furthermore, the corresponding closest incoherent state is $\sigma_{\rho}=\frac{1}{2}I_2$.

(1) Due to the spectrum decomposition of $\rho$, one has

\begin{align}
 \sqrt{\rho}=|\cos\theta|\ket{\phi_1}\bra{\phi_1}+|\sin\theta|\ket{\phi_2}\bra{\phi_2},
 \end{align}
 with $\ket{\phi_{1(2)}}=(\frac{1}{\sqrt{2}},\pm\frac{1}{\sqrt{2}})^t$. Therefore, these two sates to be discriminated are
 \begin{align}
&\ket{\psi^{\prime}_1}=\frac{1}{\sqrt{2}}(|\cos\theta|+|\sin\theta|,|\cos\theta|-|\sin\theta|)^t,\nonumber\\
&\ket{\psi^{\prime}_2}=\frac{1}{\sqrt{2}}(|\cos\theta|-|\sin\theta|,|\cos\theta|+|\sin\theta|)^t.\nonumber
\end{align}

(2) Without of generality, assuming $\theta\in[0,\frac{\pi}{2}]$, then the unitary, that maps $\ket{\psi^{\prime}_i}$ to $\ket{\psi_i}$ for $i=1,2$, is
\begin{align}
U=\begin{pmatrix}
\frac{1}{\sqrt{2}}&\frac{1}{\sqrt{2}}\\
\frac{1}{\sqrt{2}}&-\frac{1}{\sqrt{2}}
\end{pmatrix}.\nonumber
\end{align}

(3) As the components of $\{\ket{\psi^{\prime}_1},\ket{\psi^{\prime}_2}\}$ are all real, we can assume the optimal measurement $\{\ket{f^{\prime}_1},\ket{f^{\prime}_2}\}$ as
\begin{align}
&\ket{f^{\prime}_1}=\cos\vartheta\ket{1}+\sin\vartheta \ket{2},\nonumber\\
&\ket{f^{\prime}_2}=-\sin\vartheta\ket{1}+\cos\vartheta \ket{2}.\nonumber
\end{align}

Based on (\ref{eq1}), we have that $\cos\vartheta=1$.

(4) As a result, the optimal measurement to discriminate $\{\ket{\psi_i},\frac{1}{2}\}$ is
\begin{align}
&\ket{f_1}=U\ket{f^{\prime}_1}=(\frac{1}{\sqrt{2}},\frac{1}{\sqrt{2}})^t,\nonumber\\
&\ket{f_2}=U\ket{f^{\prime}_2}=(\frac{1}{\sqrt{2}},-\frac{1}{\sqrt{2}})^t.\nonumber
\end{align}
This result is consistent with (\ref{eq14}).

\section{conclusion}

In this paper, we reveal the equivalence between geometric coherence and the task to discriminate a set of pure states. This equivalence provides an operational interpretation for geometric coherence. Based on the equivalence, we prove a tighter upper bound for geometric coherence and obtain an explicit expression for the geometric coherence of generalized X-states. Moreover, we reveal the duality relationship between geometric coherence and path distinguishability, with which a relationship between geometric coherence and $l_1$-norm of coherence is obtained.  Finally, we obtain the minimum error probability and the optimal measurement for the discrimination task of two linearly independent pure states through calculating the geometric coherence of corresponding QSD-state and CIS.

Our results not only relate the coherence theory to quantum state discrimination, but also obtain many results based on this relationship. Besides, our results may shed new light on the problem of distinguishing three linearly independent pure states.

\begin{acknowledgments}
The authors thank Professor A.K.Pati for useful comments. The project is supported by National Natural Science Foundation of China (Nos.11171301, 11571307).
\end{acknowledgments}

\appendix

\section{proof of Proposition \ref{prop1}}\label{app1}

To prove that $P^{opt }_S(\{\ket{\psi_i},\eta_i\}^d_{i=1})=P^{opt }_S(\{\ket{\phi_i},\eta_i\}^d_{i=1})$, the following lemma is helpful.
\begin{lem}\label{lem2}
Let $\{\ket{\psi_i}\}^d_{i=1}$ and $\{\ket{\phi_i}\}^d_{i=1}$ are two group of unit vectors in $\mathcal{H}$, and $\langle\psi_i|\psi_j\rangle=\langle\phi_i|\phi_j\rangle$ for all $i,j$. Then, there exist a unitary matrix $U\in U(\mathcal{H})$, such that $\ket{\psi_i}=U\ket{\phi_i}$ for each $i$.
\end{lem}
\begin{proof}
Defining matrix $\psi=(\ket{\psi_1},...,\ket{\psi_d})$, namely $\psi_{ij}$ is the $i$-th entry of $\ket{\psi_j}$. Similarly, we can also define a matrix $\phi=(\ket{\phi_1},...,\ket{\phi_d})$.

If $\langle\psi_i|\psi_j\rangle=\langle\phi_i|\phi_j\rangle$, one has $\psi^{\dagger}\psi=\phi^{\dagger}\phi$ and the $(i,j)$-th entry of $\psi^{\dagger}\psi$ is $\langle\psi_i|\psi_j\rangle$. Denoting $\psi^{\dagger}\psi=\phi^{\dagger}\phi=|A|^2$, polar decomposition theorem ensures that there exist two unitary matrices $W$ and $V$, such that $\psi=W|A|$ and $\phi=V|A|$. Consequently, one has
\begin{align}
\psi=U\phi,U=WV^{\dagger},\nonumber
\end{align}
that is,
\begin{align}
(\ket{\psi_1},...,\ket{\psi_d})=(U\ket{\phi_1},...,U\ket{\phi_d}).\nonumber
\end{align}
\end{proof}

Therefore, one has
\begin{align}
P^{opt}_S(\{\ket{\psi_i},\eta_i\}^d_{i=1})&=\max_{\{M_i\}^d_{i=1}}\sum_i\eta_i\mathrm{Tr}(M_i\ket{\psi_i}\bra{\psi_i})\nonumber\\
&=\max_{\{M_i\}^d_{i=1}}\sum_i\eta_i\bra{\phi_i}U^{\dagger}M_iU\ket{\phi_i}\nonumber\\
&=\max_{\{N_i\}^d_{i=1}}\sum_i\eta_i\mathrm{Tr}(N_i\ket{\phi_i}\bra{\phi_i})\nonumber\\
&=P^{opt}_S(\{\ket{\phi_i},\eta_i\}^d_{i=1}).\nonumber
\end{align}
The last $"="$ is the fact that if $\{M_i\}^d_{i=1}$ is a POVM in $\mathcal{H}$, so is $\{U^{\dagger}M_iU\}^d_{i=1}$. Therefore, $\{U^{\dagger}M^{opt}_iU_i\}^d_{i=1}$ is an optimal measurement for QSD task $\{\ket{\phi_i},\eta_i\}^d_{i=1}$ when $\{M^{opt}_i\}^d_{i=1}$ is optimal to discriminate $\{\ket{\psi_i},\eta_i\}^d_{i=1}$.

If $\langle\psi_i|\psi_j\rangle=\langle\phi_j|\phi_i\rangle$, that is,
$\psi^{\dagger}\psi=\overline{\phi^{\dagger}\phi}$. For Lemma 8, there exist a unitary $U^{\prime}$ such that
$\ket{\psi_i}=U^{\prime}\overline{\ket{\phi_i}}$. Similarly, we can have the same result.



%

\end{document}